\documentclass[twoside]{dis07}
\usepackage[latin1]{inputenc}
\usepackage[dvips]{graphicx,epsfig,color}
\usepackage{wrapfig,rotating}
\usepackage{amssymb,amsmath,array}

\begin{document}
\title{A Global Fit to Scattering Data with NLL BFKL Resummations}

\author{Chris White$^1$ \thanks{Supported by the Dutch Foundation for Fundamental Research on Matter (FOM).} $\,$ and Robert Thorne$^2$ \thanks{Royal Society University Research Fellow}
%
\vspace{.3cm}\\
%
1- NIKHEF, Kruislaan 409, Amsterdam 1098SJ - The Netherlands
%
\vspace{.1cm}\\
2- Department of Physics and Astronomy, University College London, WC1E 6BT, UK
}

\maketitle

\begin{abstract}
We perform a global parton fit to DIS and related data, including next-to-leading logarithmic (NLL) BFKL resummations in both the massless and massive sectors. The resummed fit improves over a standard next-to-leading order (NLO) DGLAP fit, with a positive definite gluon at the input scale as opposed to the negative gluon seen at NLO. Furthermore, the predicted longitudinal structure function is free of perturbative instability at small $x$, and the reduced cross-section shows a turnover at high $y$ (absent in the NLO fit) consistent with the HERA data. 
\end{abstract}

\section{Small $x$ Resummation}
\subsection{Motivation}
Current and forthcoming particle collider experiments involve very high energies, such that the momentum fractions $x$ of initial state partons are extremely small. The splitting functions that govern the evolution of parton densities $f_i(x,Q^2)$ with momentum scale $Q^2$, together with the coefficients that relate these partons to proton structure functions, are unstable at low Bjorken $x$ values due to terms behaving like $x^{-1}\alpha_S^n\log^m(1/x)$ where $n\geq m+1$. Although the standard DGLAP theory (where the splitting and coefficient functions are considered at a fixed order in $\alpha_S$) works well in parton fits, there is some evidence that a resummation of small $x$ logarithms is necessary. Previous work has shown that a LL analysis fails to describe data well. One resums small $x$ logarithms in the gluon density by solving the \emph{BFKL equation}~\cite{BFKL}, an integral equation for the unintegrated gluon 4-point function. One then relates this gluon to structure functions using the $k_T$ factorisation formalism~\cite{CollinskT,CatanikT} to obtain the resummed splitting and coefficient functions. 
\subsection{Solution of the BFKL equation}
Introducing the double Mellin transformed unintegrated gluon density:
\begin{equation}
f(\gamma,N)=\int_0^\infty (k^2)^{-\gamma-1}\int_0^1 dx x^N f(x,k^2),	
\label{Mellin}
\end{equation}
the NLL BFKL equation in $(N,\gamma)$ space is a double differential equation in $\gamma$:
\begin{align*}
\frac{d^2f(\gamma,N)}{d\gamma^2}&=\frac{d^2f_I(\gamma,Q_0^2)}{d\gamma^2}-\frac{1}{\bar{\beta}_0 N}\frac{d(\chi_0(\gamma)f(\gamma,N))}{d\gamma}\notag\\
&+\frac{\pi}{3\bar{\beta}_0^2 N}\chi_1(\gamma)f(\gamma,N),
\end{align*}
with $\bar{\beta}_0=3/(\pi\beta_0)$. The derivatives in $\gamma$ arise from the use of the LO running coupling $\alpha_S(k^2)=1/(\beta_0\log{k^2/\Lambda^2})$ in momentum space, and $\chi_n(\gamma)$ is the Mellin transform of the $n^{\text{th}}$-order BFKL kernel. One may solve this to give:
\begin{equation}
f(N,\gamma)=\exp\left(-\frac{X_1(\gamma)}{\bar{\beta}_0 N}\right)\int_\gamma^\infty A(\tilde{\gamma})\exp\left(\frac{X_1(\tilde{\gamma})}{\bar{\beta}_0 N}\right)d\tilde{\gamma}
\label{sol}
\end{equation}
for some $A(\tilde{\gamma})$ and $X_1(\tilde{\gamma})$. One would ideally like to factorise the perturbative from the non-perturbative physics to make contact with the collinear factorisation framework. This can be achieved (up to power-suppressed corrections) by shifting the lower limit of the integral in equation (\ref{sol}) from $\gamma\rightarrow0$. Then one finds for the integrated gluon:
\begin{equation}
{\cal G}(N,t)={\cal G}_E(N,t){\cal G}_I(Q_0^2,N),
\end{equation}
where the perturbative piece is:
\begin{equation}
{\cal G}_E^1(N,t)=\frac{1}{2\pi\imath}\int_{1/2-\imath\infty}^{1/2+\imath\infty}\frac{f^{\beta_0}}{\gamma}\exp\left[\gamma t-X_1(\gamma,N)/(\bar{\beta}_0 N)\right]d\gamma,
\end{equation}
where $X_1$ can be derived from $\chi_0(\gamma)$ and $\chi_1(\gamma)$, and $f^{\beta_0}$ is a known function of $\gamma$. Structure functions have a similar form:
\begin{equation}
{\cal F}_E^1(N,t)=\frac{1}{2\pi\imath}\int_{1/2-\imath\infty}^{1/2+\imath\infty}\frac{h(\gamma,N)f^{\beta_0}}{\gamma}\exp\left[\gamma t-X_1(\gamma,N)/(\bar{\beta}_0 N)\right]d\gamma,
\end{equation}
where $h(\gamma,N)$ is a NLL order impact factor coupling the virtual photon with the BFKL gluon. If all impact factors are known, one can derive all necessary splitting and coefficient functions in double Mellin space (within a particular factorisation scheme) by taking ratios of the above quantities. The non-perturbative dependence then cancels, and one obtains results in momentum and $x$ space by performing the inverse Mellin integrals either numerically or analytically. The exact NLL impact factors are not in fact known, but the LL results supplemented with the correct kinematic behaviour of the gluon have been calculated~\cite{Peschanski,WPT}. We have shown that one expects them to approximate well the missing NLL information in the true impact factors~\cite{WT1}. \\
Consistent implementation of small $x$ resummations in the massive sector requires the definition of a variable flavour number scheme that allows the massive impact factors to be disentangled in terms of heavy coefficient functions and matrix elements. We have devised such a scheme, the DIS($\chi$) scheme~\cite{WT2}. With resummations in both the massive and massless sectors, one has everything necessary to carry out a global fit to DIS and related data. First, the resummed splitting and coefficient functions are combined with the NLO DGLAP results using the prescription:
\begin{equation*}
P^{tot.}=P^{NLL}+P^{NLO}-\left[P^{NLL(0)}+P^{NLL(1)}\right],
\end{equation*}
where the subtractions remove the double counted terms, namely the LO and NLO (in $\alpha_S$) parts of the resummed results. Then the resulting improved splitting and coefficient functions interpolate between the resummed results at low $x$, and the conventional DGLAP results at high $x$.
\section{Results}
\begin{wrapfigure}{}{0.6\columnwidth}
\centerline{\includegraphics[width=0.6\columnwidth]{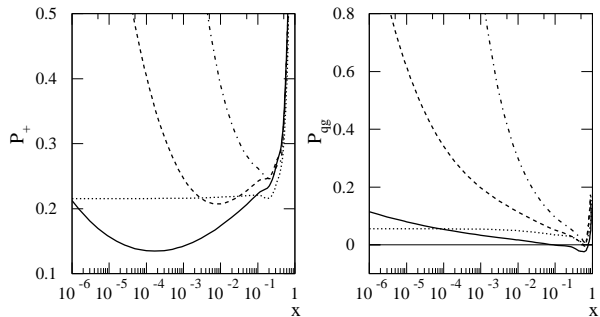}}
\caption{Splitting functions in the DIS scheme for $n_f=4$, $t=\log(Q^2/\Lambda^2)=6$: NLL+NLO (solid); LL with running coupling + LO (dashed); LL + LO (dot-dashed); NLO (dotted).}\label{pgg}
\end{wrapfigure}

\begin{figure}[t!]
\begin{center}
\scalebox{0.5}{\includegraphics{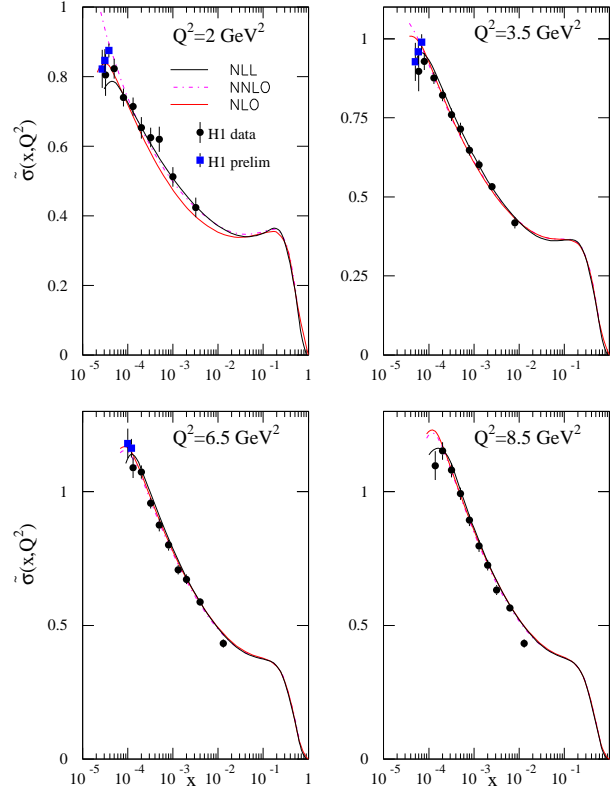}}
\caption{Reduced cross-section data, compared with both resummed and fixed order predictions.}\label{redxsec}
\end{center}
\end{figure}

The resummed splitting functions $P_{+}$ ($\simeq P_{gg}+4/9 P_{qg}$ at small $x$) and $P_{qg}$ are shown in figure \ref{pgg}. One sees that the LL BFKL results are much more divergent than the standard NLO results, which are known to describe data well. The addition of the running coupling to the LL BFKL equation suppresses this divergence, but it is still unacceptable. Inclusion of the NLL BFKL kernel, however, leads to a significant dip of the splitting functions below the NLO results. This dip is also observed in other resummation approaches~\cite{ABF,CCSS} and has an important consequence in the global fit in that it resolves the tension between the Tevatron jet data (which favour a larger high $x$ gluon) and the H1 and ZEUS data (which prefer a higher low $x$ gluon). By momentum conservation, one cannot increase the gluon at both low and high $x$ in a standard NLO DGLAP fit. This is possible in the resummed approach, due to the dip in the splitting functions.  \\
Indeed, the gluon distribution at the parton input scale of $Q_0^2=1\text{GeV}^2$ is positive definite over the entire $x$ range. This is in contrast to a NLO fit, where the gluon distribution is negative at small $x$ for low $Q^2$ values. Whilst a negative gluon is not disallowed, it can lead to negative structure functions which are unphysical. The resummed gluon, however, leads to a prediction for the longitudinal structure function that is positive and growing at small $x$ and $Q^2$, in contrast to fixed order results which show a significant perturbative instability.\\
A consequence of a more sensible description for $F_L$ is that a turnover is observed in the reduced cross-section $\tilde{\sigma}=F_2-y^2/[1+(1-y)^2]\,F_L$ at high $y$. As seen in figure \ref{redxsec}, this is required by the HERA data. Furthermore, this feature is missing in NLO fits (but present at NNLO). Thus the resummations lead to qualitatively different behaviour, consistent with known consequences of higher orders in the fixed order expansion. Overall, we find very compelling evidence of the need for BFKL effects in describing proton structure \cite{WT3}.


\begin{footnotesize}


\end{footnotesize}


\end{document}